\documentclass[amsmath,10pt,aps,prl,twocolumn,showpacs,superscriptaddress,groupedaddress]{revtex4-1}%
\usepackage{epsfig,dsfont,amssymb,amsmath,amsthm,amsfonts,amsbsy,mathrsfs}
\usepackage{graphicx}
\usepackage{amsmath}
\usepackage{amssymb}%
\usepackage{bm}
\usepackage{multirow}
\usepackage{times}

\usepackage{color}

\usepackage{hyperref}
\hypersetup{  colorlinks=true, linkcolor=blue, citecolor=red, urlcolor=blue  }

\usepackage{physics}

\theoremstyle{definition}

\newtheorem{theorem}{Theorem}

\begin{document}

\title{Universal Bound on Sampling Bosons in Linear Optics}

\author{Man-Hong Yung}
\email{yung@sustc.edu.cn}
\affiliation{Department of Physics, South University of Science and Technology of China, Shenzhen, Guangdong 518055, China}

\author{Xun Gao}
\affiliation{Center for Quantum Information, Institute for Interdisciplinary Information Sciences, Tsinghua University, Beijing 100084, China} 

\author{Joonsuk Huh}
\email{huhsh@postech.ac.kr}
\affiliation{Mueunjae Institute for Chemistry (MIC), Department of Chemistry, Pohang University of Science and Technology (POSTECH), Pohang 790-784, Korea}


\begin{abstract}
    In linear optics, photons are scattered in a network through passive optical elements including beamsplitters and phase shifters, leading to many intriguing applications in physics, such as Mach-Zehnder interferometry, Hong-Ou-Mandel effect, and tests of fundamental quantum mechanics. Here we present a general analytic expression governing the upper limit of the transition amplitudes in sampling bosons, through all realizable linear optics. Apart from boson sampling, this transition bound results in many other interesting applications, including behaviors of Bose-Einstein Condensates (BEC) in optical networks, counterparts of Hong-Ou-Mandel effects for multiple photons, and approximating permanents of matrices. Also, this general bound implies the existence of a polynomial-time randomized algorithm for estimating transition amplitudes of bosons, which represents a solution to an open problem raised by Aaronson and Hance in 2012.
\end{abstract}
\maketitle

{\bf Introduction.} Apart from being of fundamental interest in physics, linear optics has become a simple but powerful tool for processing quantum information~\cite{nielsen2011quantum,Ox2008,Peruzzo2013,Chapman2016a} and quantum simulation~\cite{Lu2009,Lanyon2010,Aspuru-Guzik2012a,Xu2012}. One of the major advantages for encoding information with light is that photons are highly robust against decoherence, which makes it an ideal system to study quantum coherence~\cite{scully1997quantum,Xu2013,Tillmann2015,Biggerstaff2016}.Furthermore, in linear optical networks, all possible transformations can be achieved with simple operations involving at most a pair of modes; more precisely, every optical circuit can be implemented with beamsplitters and phase shifters only~\cite{Reck1994}. Linear optical networks have been routinely built in photonic chips~\cite{Crespi2011,Shadbolt2012,Carolan2015} using standard semiconductor fabrication technology. In particular, a re-programmable linear optical circuit has been integrated into a photonic chip~\cite{Carolan2015}, which can perform universal operations on six photonic modes with up to six photons.

Recently, it was found that boson sampling~\cite{Aaronson2013a}, as a novel application of linear optics, can be regarded as evidence for proving the inefficiency of classical devices to perform quantum simulation, which represents a serious challenge to the validity of the extended Church-Turing thesis~\cite{nielsen2011quantum}. In boson sampling~\cite{Aaronson2013a}, a product of single-photon states is injected into a linear photonic network that encodes an instance of complex matrices. The ability of simulating boson sampling with classical devices implies the ability to approximate the corresponding permanents of matrices with a multiplicative error, which is widely believed to be impossible, based on computational complexity assumptions~\cite{Aaronson2013a}. With this motivation, much progress~\cite{Broome2013,Spring2013,Tillmann2013,Crespi2013,Spagnolo2014,Bentivegna2015,Carolan2015} has been made in the experimental realization of boson samplers using linear optical devices.

Here we present a theoretical bound on the transition amplitudes for sampling bosons, from a product of Fock states to another, through linear optics. This bound limits the efficiency of sampling bosons for all possible linear optical networks, including the the behaviors of Bose-Einstein Condensates (BEC) in optical networks, and the counterparts of Hong-Ou-Mandel effects for multiple photons.

Furthermore, our bound is important for our proof on the existence of a polynomial-time randomized algorithm for approximating permanents of matrices; this result resolves an open problem~\cite{Aaronson2012b} raised by Aaronson and Hance. 

The open question~\cite{Aaronson2012b} is: ``can we estimate any linear-optical amplitude (see Eq.~(\ref{amplitude_perm})) to $\pm 1/{\text{poly}}(n)$ additive error (or better) in polynomial time?". For the special case where the initial state is the same as that in Boson Sampling, i.e., $\left| {{t_1}{t_2}...{t_m}} \right\rangle  = \left| {111..00} \right\rangle $. Aaronson and Hance confirmed that such algorithm exists through a modification of the Gurvits algorithm. 

With our bound shown in Eq.~(\ref{general_bound}), we confirm that there exists a polynomial-time randomized algorithm for the general cases. Thus, the open problem is now completely settled.

{\bf Transition amplitude in linear optics.} The problem of interest in this work is described as follows: let us suppose that we are given a product of Fock states containing a total of $n$ identical photons (or generally bosons) distributed over $m$ different modes, i.e., $\left| {{t_1}{t_2} \cdots {t_m}} \right\rangle  \equiv \left| {{t_1}} \right\rangle  \otimes \left| {{t_2}} \right\rangle  \otimes \cdots \otimes \left| {{t_m}} \right\rangle $, where $\left| {{t_k}} \right\rangle  \equiv {({t_k}!)^{ - 1/2}}{(a_k^\dag )^{{t_k}}}\left| {{\text{vac}}} \right\rangle$ contains $t_k$ photons for $t_k = 0, 1, 2,...,n$. 

Moreover, the state is subject to the constraint of particle conservation: $\sum\nolimits_{k = 1}^m {{t_k}}  = n$. Here $a_k^{\dagger}$ creates a boson in $k$-th mode and satisfies the commutation relations: $[{a_j},a_k^\dag ] \equiv {a_j}a_k^\dag  - a_k^\dag {a_j} = {\delta _{ij}}$, $[a_j^\dag ,a_k^\dag ] = [{a_j},{a_k}] = 0$. 

Let us consider any member $U$ in the set of all possible unitary operators (i.e., linear optics) that induces a linear transformation (i.e., non-interacting) for the boson modes, i.e., 
\begin{equation}\label{U_transform}
Ua_k^\dag {U^\dag } = \sum\limits_{j = 1}^m {{u_{kj}}} a_j^\dag \ .
\end{equation}
The central problem is to give an upper bound for the absolute value of the transition amplitude, $ | \left\langle {{s_1}{s_2}\cdots{s_m}} \right|U\left| {{t_1}{t_2}\cdots{t_m}} \right\rangle |$, for locating the resulting state in another given product state, $\left| {{s_1}{s_2} \cdots {s_m}} \right\rangle$, subject to the same particle-conserving constraint $\sum\nolimits_{k = 1}^m {{s_k}}  = n$, for $s_k = 0, 1, 2,...,n$.

{\bf Main result.} Here we shall prove that the upper bound of the boson transition amplitude is given by the following expression:
\begin{equation}\label{general_bound}
\left| {\langle {s_1} \cdots {s_m}|U\left| {{t_1} \cdots {t_m}} \right\rangle } \right| \leqslant \min \left\{ {{v_{\bm s}}/{v_{\bm t}},{v_{\bm t}}/{v_{\bm s}}} \right\} \ ,
\end{equation}
where $v_{\bm s}$ is a product of $m$ factors generated from the elements in the list ${\bm s}=(s_1,s_2,...,s_m) $, 
\begin{equation}\label{define_vs}
{v_{\bm s}} \equiv \sqrt {({s_1}!/s_1^{{s_1}})({s_2}!/s_2^{{s_2}}) \cdots ({s_m}!/s_m^{{s_m}})} \ ,
\end{equation}
and defined similarly for $v_{\bm t}$. If one of the modes is unoccupied, e.g., $s_k =0$, then we simply set ${s_k}!/s_k^{{s_k}} \to 1$. 

In the context of boson sampling~\cite{Aaronson2013a}, the initial state is always a product of single-photon states, i.e., $\left| {{t_1}{t_2} \cdots {t_m}} \right\rangle  = \left| {111 \cdots 00} \right\rangle$. In this case, we can recover the result obtained previously by Aaronson and Hance~\cite{Aaronson2012b}, i.e., $\left| {\left\langle {{s_1}{s_2} \cdots {s_m}} \right|U\left| {111 \cdots 00} \right\rangle } \right| \leqslant {v_{\bm s}}$.

{\bf Direct applications.} As an application, the bound found by Aaronson and Hance implies that~\cite{Aaronson2012b} for the case where ${s_1} = n$, and ${s_2} = {s_3} \cdots  = {s_m} = 0$, the probability of putting all bosons into the same mode from $\left| {111..00} \right\rangle$ is exponentially low, as 
\begin{equation}\label{exp_low}
{P_{\max }}\left( {n,0...0|1,1...1} \right) = {v_{\bm s}^2} = {n!/{n^n}}  \approx \sqrt {2\pi n} \, {e^{ - n}} \ ,    
\end{equation}
using the Stirling approximation, $n! \approx \sqrt {2\pi n} \, {\left( {n/e} \right)^n}$. 

Consequently, for $n \geq 3$, one cannot observe the counterpart of Hong-Ou-Mandel dip with linear optics~\cite{Aaronson2012b}; the reason why Hong-Ou-Mandel dip is possible for the case of $n=2$ is because the bound is given by ${P_{\max }}\left( {2,0|1,1} \right) = v_{\bm s}^2 = 2!/{2^2} = 1/2 $, but there are two modes; so the total probability can reach unity. 

With the more general bound shown in Eq.~(\ref{general_bound}), we can bound the transition probabilities for more scenarios. For example, imagine there are $p$ bosons in one mode and $q $ bosons in another mode. The probability of getting $p+q$ in a single mode through linear optics is then bounded by the following, 
\begin{equation}\label{p_q0pq}
{P_{\max }}\left( {p + q,0|p,q} \right) = \frac{{(p + q)!}}{{p!q!}}\frac{{{p^p}{q^q}}}{{{{(p + q)}^{p + q}}}} \ ,
\end{equation}
or its inverse. We can, for instance, ask the following questions. 

Question 1: {\it can we apply linear optics to create a mode with $2n$ bosons from two separate modes with $n$ bosons each?} Unlikely. In this case, ${P_{\max }}\left( {2n|n,n} \right) = 2n!/n{!^2}{2^{2n}}$. In the limit of Bose-Einstein Condensate (BEC), where $n \gg 1$,  the probability bound, ${P_{\max }}\left( {2n|n,n} \right) \approx 1/\sqrt {\pi n}$, decreases as $O(1/\sqrt{n})$. An optimal strategy for achieving the bound is to apply the 50:50 beamsplitter, i.e., $a_1^\dag  \to (a_1^\dag  + a_2^\dag )/\sqrt 2 $ and $a_2^\dag  \to (a_1^\dag  - a_2^\dag )/\sqrt 2$.

Question 2: {\it can we add $1$ extra boson to a BEC using linear optics?} Yes! Suppose $p=n$ and $q=1$, the bound is given by ${P_{\max }}\left( { n+1,0|n,1} \right) = {\left( {n/(n + 1)} \right)^n}$, which approaches a constant limit when $n \to \infty$. In fact, this bound can be saturated by the following transformation: $U_n a_1^\dag {U_n^\dag } = \cos \theta_n \, a_1^\dag  + \sin \theta_n \, a_2^\dag$, where ${\sin ^2}\theta_n  = {(n + 1)^{ - 1}}$. We provide justification on the validity of the optimal strategies for both questions in the appendix.

{\bf Connection with permanents of matrices.} Another implication of our main result is related to permanents of matrices. The transition amplitude in Eq.~(\ref{general_bound}) is known (see e.g.~\cite{Aaronson2013a}) to be related to a permanent of a matrix regarding the unitary operator $U$:
\begin{equation}\label{amplitude_perm}
\left\langle {{s_1}{s_2}...{s_m}} \right|U\left| {{t_1}{t_2}...{t_m}} \right\rangle  = \frac{{{\text{Perm}}\left( {{U_{{\bm s},{\bm t}}}} \right)}}{{\sqrt {{s_1}! \cdots {s_m}! \cdot {t_1}! \cdots {t_m}!} }} \ ,
\end{equation}
where $U_{{\bm s},{\bm t}}$ is a $n \times n$ matrix constructed by the transformation elements $u_{kj}$ (see Eq.~(\ref{U_transform})) of the unitary operator~$U$ in the following way: create $s_k$ copies of a row vectors ${\bm \mu}_{k,{\bm t}}$ that contains $t_j$ copies of $u_{kj}$'s. For example, if ${\bm s} = \left( {1,0,2} \right)$, and ${\bm t} = \left( {2,1,0} \right)$, then the matrix $U_{{\bm s},{\bm t}}$ is of the following form:                                                                                                                                                                                       
\begin{equation}
{U_{{\bm s},{\bm t}}} = \left[ {\begin{array}{*{20}{c}}
  {{{\bm \mu} _{1,{\bm t}}}} \\ 
  {{{\bm \mu} _{3,{\bm t}}}} \\ 
  {{{\bm \mu} _{3,{\bm t}}}} 
\end{array}} \right] = \left[ {\begin{array}{*{20}{c}}
  {{u_{1,1}}}&{{u_{1,1}}}&{{u_{1,2}}} \\ 
  {{u_{3,1}}}&{{u_{3,1}}}&{{u_{3,2}}} \\ 
  {{u_{3,1}}}&{{u_{3,1}}}&{{u_{3,2}}} 
\end{array}} \right] \ .
\end{equation}
Note that if all $s$'s and $t$'s equal unity, then the transition probability is exactly the same as the permanent of the matrix defined in Eq.~(\ref{U_transform}), i.e., $\langle 11...1|U\left| {11...1} \right\rangle  = {\text{Perm}}(u_{kj}) $. Therefore, our bound also implies an upper bound of the permanent of a matrix: 
\begin{equation}\label{perm_abs_min_bound}
| \, {\text{Perm}}({U_{{\bm s},{\bm t}}}) \, | \leqslant \min \left\{ {\frac{{{v_{\bm s}}}}{{{v_{\bm t}}}},\frac{{{v_{\bm t}}}}{{{v_{\bm s}}}}} \right\} \times \prod\limits_{k = 1}^m {\sqrt {{s_k}! \, {t_k}!} } \ .
\end{equation}

Before we present the proof of the bound, let us first establish a general theorem that is crucial for our result: 
\begin{theorem}
Given any polynomial function, $f(a_1^\dag ,a_2^\dag ,...,a_m^\dag )$, of multi-mode creation operators $a_k^\dag$'s, the vacuum-to-vacuum transition amplitude (unnormalized), 
\begin{equation}\label{vac_2_vac_am}
{F_{\bm s}} \equiv \left\langle {\rm vac} \right|a_1^{{s_1}}a_2^{{s_2}} \cdots a_m^{{s_m}} \, f(a_1^\dag ,a_2^\dag ,...,a_m^\dag )\left| {\rm vac} \right\rangle \ ,    
\end{equation}
can always be expressed as a sum involving a set of weighted complex roots of unity, by mapping the boson operator, ${a_k^\dagger} \to {z_k}$, to a complex number $z_k$, and similarly $a_k \to {{\bar z}_k}$ to its complex conjugate ${\bar z}_k$:
\begin{equation}\label{Fs_sum}
{F_{\bm s}} = \frac{v_{\bm s}^2}{{{d^m}}}\sum\limits_{\{{\bm z}\}} {{\bar z}_1^{{s}_1}} {{\bar z}_2^{{s_2}} \cdots {\bar z}_m^{{s_m}} \, f({{ z}_1},{{ z}_2},...,{{z}_m})} \ ,
\end{equation}
where ${z_j} \in   \{ \sqrt {{s_j}} \, { \omega ^0}, \sqrt {{s_j}} \, {\omega ^1},..., \sqrt {{s_j}} \, {\omega ^{d - 1}}\}$ is related to one of the complex roots of unity $\omega  \equiv {e^{ - 2\pi i/d}}$, weighted by a factor $\sqrt {{s_j}} $. Here $d$ is chosen to be an integer larger than the degree of the function and the sum $\sum\nolimits_{k = 1}^m {{s_k}}$. 
\end{theorem}

Alternatively, we can write $F_{\bm s}$ in the form of an expectation value: 
\begin{equation}
    {F_{\bm s}} = {v_{\bm s}^2} \, {\mathbb{E}}[\bar z_1^{{s_1}}\bar z_2^{{s_2}} \cdots \bar z_m^{{s_m}}f({z_1},{z_2},..,{z_m})] \ ,
\end{equation}
which allows us to devise a sampling algorithm to estimate its value, as we shall discuss later. 
\begin{proof}[\bf Proof of Theorem 1]
Since all the terms in the function $f(a_1^\dag ,a_2^\dag ,...,a_m^\dag ) $ commute with one another, we can, for example, sort out the first creation operator $a_1^{\dag}$ as if it was just a real number, and write
\begin{equation}
f(a_1^\dag ,a_2^\dag ,...,a_m^\dag ) = \sum\limits_{k = 0}^d {a_1^{\dag k}{\phi _k}(a_2^\dag ,...,a_m^\dag )} \ ,    
\end{equation}
where ${{\phi _k}(a_2^\dag ,...,a_m^\dag )}$ is a resulting polynomial function without $a_1^\dagger$. Consequently, we have ${F_s} = \sum\nolimits_{k = 0}^d {\left\langle {\text{0}} \right|a_1^{{s_1}}} a_1^{\dag k}\left| {\text{0}} \right\rangle \left\langle {{\text{vac}}} \right|a_2^{{s_2}} \cdots a_m^{{s_m}} \, {\phi _k}(a_2^\dag ,...,a_m^\dag )\left| {{\text{vac}}} \right\rangle $, but there is only one non-zero term in the summation, as $\left\langle {\text{0}} \right|a_1^{{s_1}}a_1^{\dag k}\left| {\text{0}} \right\rangle  = s_1! \, {\delta _{{s_1}k}}$. 

Now, since the Kronecker delta function can be expressed (by the representation through discrete Fourier transform) as follows: ${\delta _{{s_1}k}} = \left( {1/d} \right)\sum\nolimits_{j = 0}^{d - 1} {{e^{ - \left( {2\pi ij/d} \right)({s_1} - k)}}}  = \left( {1/d} \right)\sum\nolimits_{j = 0}^{d - 1} {{\omega ^{j({s_1} - k)}}} $, we can therefore write the inner product (with $z_1 \in \{ \sqrt{s_1} \omega^0, \sqrt{s_1} \omega^1, ..., \sqrt{s_1} \omega^{d-1} \}$), 
\begin{equation}
\left\langle {\text{0}} \right|a_1^{{s_1}}a_1^{\dag k}\left| {\text{0}} \right\rangle  = \frac{s_1!}{s_1^{s_1}} \frac{1}{d}\sum\limits_{\{ {z_1}\} } {{\bar z}_1^{{s_1}} z_1^k} \ ,
\end{equation}
as a sum over all values of $z_1$, which implies that ${F_s} = ({s_1}!/s_1^{{s_1}}d)\sum\nolimits_{\{ {z_1}\} } {{\bar z}_1^{{s_1}}\left\langle {\rm vac} \right|a_2^{{s_2}} \cdots a_m^{{s_m}}f({{ z}_1},a_2^\dag ,...,a_m^\dag )\left| {\rm vac} \right\rangle }$. Next, we can define a new polynomial function, $f'(a_2^\dag ,...,a_m^\dag ) \equiv \left( {s_1!/s_1^{s_1} d} \right)\sum\nolimits_{\{ {z_1}\} } {{\bar z}_1^{{s_1}}f({{ z}_1},a_2^\dag ,...,a_m^\dag )}$, and repeat the same procedure for $a_2^{\dagger}$, and so on, which yields the result in Eq.~(\ref{Fs_sum}) at the end.
\end{proof}

{\bf Proving the main result.} We are now ready to present the proof for the bound in Eq.~(\ref{general_bound}). For this purpose, we express the transition amplitude explicitly with bonsonic operators, i.e.,
\begin{equation}\label{U_Gst}
\langle {s_1} ... {s_m}|U\left| {{t_1}...{t_m}} \right\rangle  = \frac{{G\left( {U,{\bm s},{\bm t}} \right)}}{{\sqrt {({s_1}! \cdots {s_m}!)({t_1}! \cdots {t_m}!)} }} \ ,
\end{equation}
where we defined an operator function,
\begin{equation}
    G\left( {U,{\bm s},{\bm t}} \right) \equiv \left\langle {{\text{vac}}} \right|a_1^{{s_1}} \cdots a_m^{{s_m}} \, U \, a_1^{\dag {t_1}} \cdots a_m^{\dag {t_m}}\left| {{\text{vac}}} \right\rangle \ .
\end{equation}
The proof can be completed with only three steps as follows.

{\bf Step 1 (operator-to-number conversion):} With the transformation rule given in Eq.~(\ref{U_transform}), we have $U a_1^{\dag {t_1}} \cdots a_m^{\dag {t_m}} U^\dagger = \prod\nolimits_{k = 1}^m {({u_{k,1}}} a_1^\dag  + ... + {u_{k,m}}a_m^\dag {)^{{t_k}}} $,
which is exactly a polynomial function of the creation operators. Therefore, the theorem above implies that 
\begin{equation}
G\left( {U,{\bm s},{\bm t}} \right) = \frac{{{v_{\bm s}^2}}}{{{d^m}}}\sum\limits_{\left\{ {\bm z} \right\}} (\bar z_1^{{s_1}}\bar z_2^{{s_2}} \cdots \bar z_m^{{s_m}})  \, {g\left( {\bm z} \right)} \ ,
\end{equation}
where the function $g(\bm z)$ is defined as follows: 
\begin{equation}
g\left( {\bm z} \right) \equiv \prod\nolimits_{k = 1}^m {{{({u_{k,1}}{z_1} + ... + {u_{k.m}}{z_m})}^{{t_k}}}} \ .
\end{equation}
In order to bound the absolute value of $G(U,{\bm s},{\bm t})$, it is sufficient to bound the function $g({\bm z})$ by writing its absolute value in the following form: $\left| {g\left( {\bm z} \right)} \right| = \sqrt {t_1^{{t_1}}...t_m^{{t_m}}} \prod\nolimits_{j = 1}^m {{{(1/\sqrt {{t_j}} )}^{{t_j}}}} |{u_{k,1}}{z_1} + ... + {u_{k.m}}{z_m}{|^{{t_j}}}$. 

{\bf Step 2 (arithmetic-geometric inequality):} Recall that the weighted arithmetic-geometric inequality suggests that, 
\begin{equation}
A_1^{{\lambda _1}}A_2^{{\lambda _2}} \cdots A_m^{{\lambda _m}} \leqslant {\lambda _1}{A_1} + {\lambda _2}{A_2} + ... + {\lambda _m}{A_m} \ ,    
\end{equation}
for all non-negative $A_k$'s and $\lambda_k$, subject to the constraint $\sum\nolimits_{k = 1}^m {{\lambda _k}}  = 1$. In terms of our $t$'s (by setting ${\lambda _k} = {t_k}/n$), we have ${(A_1^{{t_1}}A_2^{{t_2}} \cdots A_n^{{t_n}})^{1/2}} \leqslant {[({t_1}/n){A_1} + ({t_2}/n){A_2} +  ...  + ({t_n}/n){A_n}]^{n/2}}$. Now, let us denote ${A_j} = \left( {1/{t_j}} \right)|{u_{k,1}}{z_1} + ... + {u_{k.m}}{z_m}{|^2}$. Then, we have, 
\begin{equation}
\frac{{\left| {g\left( {\bm z} \right)} \right|}}{{\sqrt {t_1^{{t_1}}\cdots t_m^{{t_m}}} }} \leqslant {(\sum\limits_{j = 1}^m {\frac{1}{n}|{u_{k,1}}{z_1} + ... + {u_{k.m}}{z_m}{|^2}} )^{n/2}}\ .
\end{equation}

{\bf Step 3 (bounding the norms):} Note that the right-hand side is related to the $2$-norm of a vector: $\left\| {\bm z} \right\| \equiv {\left\| {\bm z} \right\|_2} = \sqrt {|{z_1}{|^2} + |{z_2}{|^2} + ... + |{z_m}{|^2}}$. To take a step further, we can always define a unitary matrix~$V$ such that ${( {V{\bm z}})_k} = {u_{k,1}}{z_1} + ... + {u_{k,m}}{z_m}$, which implies that $\left| {g\left( {\bm z} \right)} \right| \leqslant {(t_1^{{t_1}}...t_m^{{t_m}}/n^n)^{1/2}}{\left\| {V{\bm z}} \right\|^{n/2}}$. Since $\left\| V \right\| = 1$ for unitary matrices, and $\left\| {\bm z} \right\| = {({s_1} + {s_2} + ...{s_m})^{1/2}} = \sqrt n $, we have $\left| {g\left( {\bm z} \right)} \right| \leqslant {(t_1^{{t_1}}...t_m^{{t_m}})^{1/2}}$. Consequently, we have 
\begin{equation}\label{GUst_bound}
|G\left( {U,{\bm s},{\bm t}} \right) |\leqslant v_{\bm s}^2\sqrt {(s_1^{{s_1}} \cdots s_m^{{s_m}})(t_1^{{t_1}} \cdots t_m^{{t_m}})} \ ,    
\end{equation}
which implies part of the advertised result of the bound $v_{\bm s}/v_{\bm t}$ in Eq.~(\ref{general_bound}). We can repeat essentially the same procedure for the complex conjugate, $\langle {t_1} \cdots {t_m}| \ U^\dagger \left| {{s_1} \cdots {s_m}} \right\rangle$, of the transition amplitude, in order to obtain the other part, $v_{\bm s}/v_{\bm t}$. This completes our proof. $\blacksquare$ 

{\bf Permanent by Sampling.} It is known that any $m \times m$ matrix $W=(w_{i,j})$ permanent can be calculated exactly with a scaling $O({m^2} \, {2^m})$ using Ryser's formula~\cite{Ryser}. Glynn~\cite{Glynn2010,Glynn2013} suggested a different algorithm requiring a similar computational cost that the Glynn's formula is given as a normalized form: ${\text{Perm}}\left( W \right) = 2^{-m}\sum\nolimits_{\bm x} {{x_1} \cdots {x_m}\prod\nolimits_{i = 1}^m {({w_{i,1}}{x_1} + \cdots + {w_{i,m}}{x_m})} }$, summing over all possible $m$-bit strings ${\bm x} = (x_1,x_2,...,x_m) \in \{ \pm 1\}^m$, then normalizing with the total number of summations $2^m$. Based on the Glynn's formula, Gurvits~\cite{Gurvits2005} proposed a polynomial-time randomized algorithm to produce an approximation ${\text{Perm}}\left( W \right)$ to the value of the permanent, with an additive error $\pm \epsilon \, {\left\| W \right\|^m}$, i.e., $ | {\tilde {\text P}\left( W \right) - {\text{Perm}}\left( W \right)} | \leq \epsilon \, {\left\| W \right\|^m}$, where $\left\| W \right\| \equiv \mathop {\sup }\nolimits_{{\bm v} \ne 0} \left\| {W {\bm v}} \right\|/\left\| {\bm v} \right\|$. 

The main idea of Gurvits is that one can convert Glynn's formula into an expectation value: ${\text{Perm}}\left( W \right) = \mathop {\mathbb E} [{x_1} \cdots {x_m}\prod\nolimits_{i = 1}^m {({w_{i,1}}{x_1} + ... + {w_{i,n}}{x_m})} ]$, where ${\text{Gl}}{{\text{y}}}({\bm x}) \equiv {x_1} \cdots {x_m}\prod\nolimits_{i = 1}^m {({w_{i,1}}{x_1} + ... + {w_{i,m}}{x_m})}$ is Glynn's estimator~\cite{Glynn2010}. An approximation of the permanent is obtained by randomly and uniformly picking $T$ strings $\bm x_k \in \{ \pm 1\}^m$, for $k = 1,2,...,T$, and evaluate the average value: 
\begin{equation}
{\tilde {\rm P}\left( W \right)}   = \frac{1}{T} \sum\limits_{k = 1}^T {{\text{Gly}}\left( {{{\bm x}_k}} \right)}  \ .  
\end{equation}
As a result, if we take a total of $T=O(m^2/\epsilon^2)$ samples, then the Chebyshev's inequality guarantees that the failure probability, where $ | {\tilde {\text P}\left( W \right) - {\text{Perm}}\left( W \right)} | > \epsilon \, {\left\| W \right\|^m}$, can be upper-bounded with a small value.

In Ref.~\cite{Aaronson2012b}, Aaronson and Hance proposed a generalization of Gurvits's algorithm by defining a generalized Glynn's estimator, namely $ {\rm GenGly}({\bm z}) \equiv {v_{\bm s}^2} \, (\bar z_1^{{s_1}} \cdots \bar z_m^{{s_m}})\prod\nolimits_{i = 1}^m {({w_{i,1}}{z_1} +  ...  + } {w_{i,m}}{z_m})$, where $v_{\bm s}$ is defined in Eq.~(\ref{define_vs}). Sampling the generalized Glynn's estimator over the complex values, the permanent, ${\rm Perm} (V)$, of a matrix $V$, which is obtained by repeating $s_i$ times the $i$-th row of the $m \times m$ matrix $W=(w_{i,j})$, can be estimated in polynomial time with an additive error $\pm \, \epsilon \, {v_s} \sqrt {{s_1}! \cdots {s_m}!} \, {\left\| W \right\|^n}$.

{ \bf Sampling algorithm for transition amplitudes.} Comparing the right-hand sides of Eq.~(\ref{amplitude_perm}) and Eq.~(\ref{U_Gst}), we concluded that $G\left( {U,{\bm s},{\bm t}} \right)$ is equal to the permanent of the matrix $U_{{\bm s},{\bm t}}$, i.e.,
\begin{equation}\label{GUst_perm}
G\left( {U,{\bm s},{\bm t}} \right) = {\text{Perm}}\left( {{U_{{\bm s},{\bm t}}}} \right)  \ .
\end{equation}
 It is therefore possible to extend our formalism for an arbitrary $m \times m$ matrix $W=(w_{i,j})$ from the transformation in Eq.~(\ref{U_transform}), which implies that we can define an even more general Glynn estimator,
\begin{equation}\label{mGenGly}
    {\rm mGenGly}({\bm z}) \equiv v_{\bm s}^2{(\prod\limits_{k = 1}^m {\bar z_k^{{s_k}})} \prod\limits_{i = 1}^m ({\sum\limits_{j = 1}^m {{w_{i,j}}{z_j}} } )^{{t_i}}} \ ,
\end{equation}
which is reduced to the estimator, ${\rm GenGly}({\bm z})$, of Aaronson and Hance for the cases where $t_1 =t_2 = ...=t_m =1$, and further reduced to the estimator, ${\rm Gly}({\bm z})$, of Gurvits, when $s_1 =s_2 = ...=s_m =1$ in addition. An alternative estimator can be found in Huh~\cite{huh2016}.

As a result, taking a total of $T=O(1/\epsilon^2)$ samples, the error in estimating the permanent is bounded by $ \pm \epsilon {\left\| W \right\|^m}v_s^2\prod\nolimits_{k = 1}^m {{{(s_k^{{s_k}}t_k^{{t_k}})}^{1/2}}}$, derived from the bound established in Eq.~(\ref{GUst_bound}). Note that the evaluation of each sample requires $O(m^2)$ steps, as in Eq.~(\ref{mGenGly}), the calculation of the summation takes $m$ steps and there are $m$ factors to multiply.

Return to the case of quantum optics, where the transformation is necessarily a unitary matrix $U$, with $\left\| U \right\| = 1$. With the identification in Eq.~(\ref{GUst_perm}) and our bound in Eq.~(\ref{general_bound}), we can therefore approximate the transition amplitude with a high probability, by uniformly sampling the more general Glynn's estimator in Eq.~(\ref{mGenGly}), with ${{\bm z}_k} \in   \{ \sqrt {{s_j}} \, { \omega ^0}, \sqrt {{s_j}} \, {\omega ^1},..., \sqrt {{s_j}} \, {\omega ^{d - 1}}\}^m$, i.e.,
\begin{equation}\label{sUt_appr_1TGen}
\langle {s_1} \cdots {s_m}|U\left| {{t_1} \cdots {t_m}} \right\rangle  \approx \frac{1}{T} \frac{{\sum\nolimits_{k = 1}^T {{\text{mGenGly(}}{{\bm z}_k}{\text{)}}} }}{{\prod\nolimits_{k = 1}^m {{{({s_k}! \, {t_k}!)}^{1/2}}} }} \ .
\end{equation}
With a polynomial number of sampling, $T=O(1/\epsilon^2)$, the error of the approximation of the transition amplitude in Eq.~(\ref{sUt_appr_1TGen}) is bounded by $\pm \epsilon \times \min \left\{ {{v_{\bm s}}/{v_{\bm t}},{v_{\bm t}}/{v_{\bm s}}} \right\}$, from the Chebyshev's inequality (see Appendix for details). The existence of this polynomial-time  algorithm, scaling as $O(m^2/\epsilon^2)$, represents a solution to an open problem raised in the work~\cite{Aaronson2012b} of Aaronson and Hance. 

{\bf Conclusion.} We have presented a general upper bound (Eq.~(\ref{general_bound})) on the transition amplitudes in sampling bosons for any linear optical network (Eq.~(\ref{U_transform})). This bound can directly be applied to many different physical scenarios such as Hong-Ou-Mandel and BEC (see Eq.~(\ref{exp_low}) and Eq.~(\ref{p_q0pq}), respectively). The crucial step in proving the bound involves a general theorem~(see Eq.~(\ref{vac_2_vac_am})) that makes it possible to convert any vaccum-to-vaccum transition amplitude, for some polynomial functions of the boson operators, into a sum of discrete random variables~(Eq.~(\ref{Fs_sum})). In addition to boson sampling, this theorem is applicable to the calculation of elements of the $S$-matrix in quantum electrodynamics (see Ref.~\cite{Caianiello1953}).

The connection between the transition amplitudes and the permanents makes it possible to bound the absolute value of the corresponding permanents of matrices~(Eq.~(\ref{perm_abs_min_bound})). Moreover, the classical algorithm proposed by Gurvits~\cite{Gurvits2005}, extended by Aaronson and Hance~\cite{Aaronson2012b}, can be further extended~(Eq.~(\ref{sUt_appr_1TGen})) with our bound; the existence of such algorithm implies that ∂the open problem of Aaronson and Hance in Ref.~\cite{Aaronson2012b} (page 16) can now be settled. 

Finally, we note that it is straightforward to show that our bound can also be applied to generalize the de-randomizing algorithm for approximating permanents of non-negative matrices, which was discussed by Aaronson and Hance~\cite{Aaronson2012b}. 

{\it Acknowledgments---}  We thank Scott Aaronson for the valuable comments on the manuscript. M.-H.Y. acknowledges support by the National Natural Science Foundation of China under Grants No. 11405093. 
J.H. acknowledges supports by Basic Science Research Program through the National Research Foundation of Korea(NRF) funded by the Ministry of Education, Science and Technology(NRF-2015R1A6A3A04059773), the ICT R\&D program of MSIP/IITP [2015-019, Fundamental Research Toward Secure Quantum Communication] and Mueunjae Institute for Chemistry (MIC) postdoctoral fellowship.


\begin{thebibliography}{30}%
\makeatletter
\providecommand \@ifxundefined [1]{%
 \@ifx{#1\undefined}
}%
\providecommand \@ifnum [1]{%
 \ifnum #1\expandafter \@firstoftwo
 \else \expandafter \@secondoftwo
 \fi
}%
\providecommand \@ifx [1]{%
 \ifx #1\expandafter \@firstoftwo
 \else \expandafter \@secondoftwo
 \fi
}%
\providecommand \natexlab [1]{#1}%
\providecommand \enquote  [1]{``#1''}%
\providecommand \bibnamefont  [1]{#1}%
\providecommand \bibfnamefont [1]{#1}%
\providecommand \citenamefont [1]{#1}%
\providecommand \href@noop [0]{\@secondoftwo}%
\providecommand \href [0]{\begingroup \@sanitize@url \@href}%
\providecommand \@href[1]{\@@startlink{#1}\@@href}%
\providecommand \@@href[1]{\endgroup#1\@@endlink}%
\providecommand \@sanitize@url [0]{\catcode `\\12\catcode `\$12\catcode
  `\&12\catcode `\#12\catcode `\^12\catcode `\_12\catcode `\%12\relax}%
\providecommand \@@startlink[1]{}%
\providecommand \@@endlink[0]{}%
\providecommand \url  [0]{\begingroup\@sanitize@url \@url }%
\providecommand \@url [1]{\endgroup\@href {#1}{\urlprefix }}%
\providecommand \urlprefix  [0]{URL }%
\providecommand \Eprint [0]{\href }%
\providecommand \doibase [0]{http://dx.doi.org/}%
\providecommand \selectlanguage [0]{\@gobble}%
\providecommand \bibinfo  [0]{\@secondoftwo}%
\providecommand \bibfield  [0]{\@secondoftwo}%
\providecommand \translation [1]{[#1]}%
\providecommand \BibitemOpen [0]{}%
\providecommand \bibitemStop [0]{}%
\providecommand \bibitemNoStop [0]{.\EOS\space}%
\providecommand \EOS [0]{\spacefactor3000\relax}%
\providecommand \BibitemShut  [1]{\csname bibitem#1\endcsname}%
\let\auto@bib@innerbib\@empty
\bibitem [{\citenamefont {Nielsen}\ and\ \citenamefont
  {Chuang}(2011)}]{nielsen2011quantum}%
  \BibitemOpen
  \bibfield  {author} {\bibinfo {author} {\bibfnamefont {M.~A.}\ \bibnamefont
  {Nielsen}}\ and\ \bibinfo {author} {\bibfnamefont {I.~L.}\ \bibnamefont
  {Chuang}},\ }\href {http://books.google.com/books?id=-s4DEy7o-a0C} {\emph
  {\bibinfo {title} {{Quantum Computation and Quantum Information: 10th
  Anniversary Edition}}}},\ Quantum Computation and Quantum Information\
  (\bibinfo  {publisher} {Cambridge University Press},\ \bibinfo {year}
  {2011})\BibitemShut {NoStop}%
\bibitem [{\citenamefont {Kok}\ \emph {et~al.}(2007)\citenamefont {Kok},
  \citenamefont {Munro}, \citenamefont {Nemoto}, \citenamefont {Ralph},
  \citenamefont {Dowling},\ and\ \citenamefont {Milburn}}]{Ox2008}%
  \BibitemOpen
  \bibfield  {author} {\bibinfo {author} {\bibfnamefont {P.}~\bibnamefont
  {Kok}}, \bibinfo {author} {\bibfnamefont {W.~J.}\ \bibnamefont {Munro}},
  \bibinfo {author} {\bibfnamefont {K.}~\bibnamefont {Nemoto}}, \bibinfo
  {author} {\bibfnamefont {T.~C.}\ \bibnamefont {Ralph}}, \bibinfo {author}
  {\bibfnamefont {J.~P.}\ \bibnamefont {Dowling}}, \ and\ \bibinfo {author}
  {\bibfnamefont {G.~J.}\ \bibnamefont {Milburn}},\ }\href {\doibase
  10.1103/RevModPhys.79.135} {\bibfield  {journal} {\bibinfo  {journal} {Rev.
  Mod. Phys.}\ }\textbf {\bibinfo {volume} {79}},\ \bibinfo {pages} {135}
  (\bibinfo {year} {2007})}\BibitemShut {NoStop}%
\bibitem [{\citenamefont {Peruzzo}\ \emph {et~al.}(2014)\citenamefont
  {Peruzzo}, \citenamefont {McClean}, \citenamefont {Shadbolt}, \citenamefont
  {Yung}, \citenamefont {Zhou}, \citenamefont {Love}, \citenamefont
  {Aspuru-Guzik},\ and\ \citenamefont {O'Brien}}]{Peruzzo2013}%
  \BibitemOpen
  \bibfield  {author} {\bibinfo {author} {\bibfnamefont {A.}~\bibnamefont
  {Peruzzo}}, \bibinfo {author} {\bibfnamefont {J.}~\bibnamefont {McClean}},
  \bibinfo {author} {\bibfnamefont {P.}~\bibnamefont {Shadbolt}}, \bibinfo
  {author} {\bibfnamefont {M.-h.}\ \bibnamefont {Yung}}, \bibinfo {author}
  {\bibfnamefont {X.-q.}\ \bibnamefont {Zhou}}, \bibinfo {author}
  {\bibfnamefont {P.~J.}\ \bibnamefont {Love}}, \bibinfo {author}
  {\bibfnamefont {A.}~\bibnamefont {Aspuru-Guzik}}, \ and\ \bibinfo {author}
  {\bibfnamefont {J.~L.}\ \bibnamefont {O'Brien}},\ }\href {\doibase
  10.1038/ncomms5213} {\bibfield  {journal} {\bibinfo  {journal} {Nat.
  Commun.}\ }\textbf {\bibinfo {volume} {5}},\ \bibinfo {pages} {4213}
  (\bibinfo {year} {2014})}\BibitemShut {NoStop}%
\bibitem [{\citenamefont {Chapman}\ \emph {et~al.}(2016)\citenamefont
  {Chapman}, \citenamefont {Santandrea}, \citenamefont {Huang}, \citenamefont
  {Corrielli}, \citenamefont {Crespi}, \citenamefont {Yung}, \citenamefont
  {Osellame},\ and\ \citenamefont {Peruzzo}}]{Chapman2016a}%
  \BibitemOpen
  \bibfield  {author} {\bibinfo {author} {\bibfnamefont {R.~J.}\ \bibnamefont
  {Chapman}}, \bibinfo {author} {\bibfnamefont {M.}~\bibnamefont {Santandrea}},
  \bibinfo {author} {\bibfnamefont {Z.}~\bibnamefont {Huang}}, \bibinfo
  {author} {\bibfnamefont {G.}~\bibnamefont {Corrielli}}, \bibinfo {author}
  {\bibfnamefont {A.}~\bibnamefont {Crespi}}, \bibinfo {author} {\bibfnamefont
  {M.-H.}\ \bibnamefont {Yung}}, \bibinfo {author} {\bibfnamefont
  {R.}~\bibnamefont {Osellame}}, \ and\ \bibinfo {author} {\bibfnamefont
  {A.}~\bibnamefont {Peruzzo}},\ }\href {\doibase 10.1038/ncomms11339}
  {\bibfield  {journal} {\bibinfo  {journal} {Nat. Commun.}\ }\textbf {\bibinfo
  {volume} {7}},\ \bibinfo {pages} {11339} (\bibinfo {year}
  {2016})}\BibitemShut {NoStop}%
\bibitem [{\citenamefont {Lu}\ \emph {et~al.}(2009)\citenamefont {Lu},
  \citenamefont {Gao}, \citenamefont {G{\"{u}}hne}, \citenamefont {Zhou},
  \citenamefont {Chen},\ and\ \citenamefont {Pan}}]{Lu2009}%
  \BibitemOpen
  \bibfield  {author} {\bibinfo {author} {\bibfnamefont {C.-y.}\ \bibnamefont
  {Lu}}, \bibinfo {author} {\bibfnamefont {W.-b.}\ \bibnamefont {Gao}},
  \bibinfo {author} {\bibfnamefont {O.}~\bibnamefont {G{\"{u}}hne}}, \bibinfo
  {author} {\bibfnamefont {X.-q.}\ \bibnamefont {Zhou}}, \bibinfo {author}
  {\bibfnamefont {Z.-b.}\ \bibnamefont {Chen}}, \ and\ \bibinfo {author}
  {\bibfnamefont {J.-w.}\ \bibnamefont {Pan}},\ }\href {\doibase
  10.1103/PhysRevLett.102.030502} {\bibfield  {journal} {\bibinfo  {journal}
  {Phys. Rev. Lett.}\ }\textbf {\bibinfo {volume} {102}},\ \bibinfo {pages}
  {030502} (\bibinfo {year} {2009})}\BibitemShut {NoStop}%
\bibitem [{\citenamefont {Lanyon}\ \emph {et~al.}(2010)\citenamefont {Lanyon},
  \citenamefont {Whitfield}, \citenamefont {Gillett}, \citenamefont {Goggin},
  \citenamefont {Almeida}, \citenamefont {Kassal}, \citenamefont {Biamonte},
  \citenamefont {Mohseni}, \citenamefont {Powell}, \citenamefont {Barbieri},
  \citenamefont {Aspuru-Guzik},\ and\ \citenamefont {White}}]{Lanyon2010}%
  \BibitemOpen
  \bibfield  {author} {\bibinfo {author} {\bibfnamefont {B.~P.}\ \bibnamefont
  {Lanyon}}, \bibinfo {author} {\bibfnamefont {J.~D.}\ \bibnamefont
  {Whitfield}}, \bibinfo {author} {\bibfnamefont {G.~G.}\ \bibnamefont
  {Gillett}}, \bibinfo {author} {\bibfnamefont {M.~E.}\ \bibnamefont {Goggin}},
  \bibinfo {author} {\bibfnamefont {M.~P.}\ \bibnamefont {Almeida}}, \bibinfo
  {author} {\bibfnamefont {I.}~\bibnamefont {Kassal}}, \bibinfo {author}
  {\bibfnamefont {J.~D.}\ \bibnamefont {Biamonte}}, \bibinfo {author}
  {\bibfnamefont {M.}~\bibnamefont {Mohseni}}, \bibinfo {author} {\bibfnamefont
  {B.~J.}\ \bibnamefont {Powell}}, \bibinfo {author} {\bibfnamefont
  {M.}~\bibnamefont {Barbieri}}, \bibinfo {author} {\bibfnamefont
  {A.}~\bibnamefont {Aspuru-Guzik}}, \ and\ \bibinfo {author} {\bibfnamefont
  {A.~G.}\ \bibnamefont {White}},\ }\href {\doibase 10.1038/nchem.483}
  {\bibfield  {journal} {\bibinfo  {journal} {Nat. Chem.}\ }\textbf {\bibinfo
  {volume} {2}},\ \bibinfo {pages} {106} (\bibinfo {year} {2010})}\BibitemShut
  {NoStop}%
\bibitem [{\citenamefont {Aspuru-Guzik}\ and\ \citenamefont
  {Walther}(2012)}]{Aspuru-Guzik2012a}%
  \BibitemOpen
  \bibfield  {author} {\bibinfo {author} {\bibfnamefont {A.}~\bibnamefont
  {Aspuru-Guzik}}\ and\ \bibinfo {author} {\bibfnamefont {P.}~\bibnamefont
  {Walther}},\ }\href {\doibase 10.1038/nphys2253} {\bibfield  {journal}
  {\bibinfo  {journal} {Nat. Phys.}\ }\textbf {\bibinfo {volume} {8}},\
  \bibinfo {pages} {285} (\bibinfo {year} {2012})}\BibitemShut {NoStop}%
\bibitem [{\citenamefont {Xu}\ \emph {et~al.}(2014)\citenamefont {Xu},
  \citenamefont {Yung}, \citenamefont {Xu}, \citenamefont {Boixo},
  \citenamefont {Zhou}, \citenamefont {Li}, \citenamefont {Aspuru-Guzik},\ and\
  \citenamefont {Guo}}]{Xu2012}%
  \BibitemOpen
  \bibfield  {author} {\bibinfo {author} {\bibfnamefont {J.-S.}\ \bibnamefont
  {Xu}}, \bibinfo {author} {\bibfnamefont {M.-H.}\ \bibnamefont {Yung}},
  \bibinfo {author} {\bibfnamefont {X.-y.}\ \bibnamefont {Xu}}, \bibinfo
  {author} {\bibfnamefont {S.}~\bibnamefont {Boixo}}, \bibinfo {author}
  {\bibfnamefont {Z.-w.}\ \bibnamefont {Zhou}}, \bibinfo {author}
  {\bibfnamefont {C.-f.}\ \bibnamefont {Li}}, \bibinfo {author} {\bibfnamefont
  {A.}~\bibnamefont {Aspuru-Guzik}}, \ and\ \bibinfo {author} {\bibfnamefont
  {G.-C.}\ \bibnamefont {Guo}},\ }\href {\doibase 10.1038/nphoton.2013.354}
  {\bibfield  {journal} {\bibinfo  {journal} {Nat. Photonics}\ }\textbf
  {\bibinfo {volume} {8}},\ \bibinfo {pages} {113} (\bibinfo {year}
  {2014})}\BibitemShut {NoStop}%
\bibitem [{\citenamefont {Scully}\ and\ \citenamefont
  {Zubairy}(1997)}]{scully1997quantum}%
  \BibitemOpen
  \bibfield  {author} {\bibinfo {author} {\bibfnamefont {M.~O.}\ \bibnamefont
  {Scully}}\ and\ \bibinfo {author} {\bibfnamefont {M.~S.}\ \bibnamefont
  {Zubairy}},\ }\href@noop {} {\emph {\bibinfo {title} {{Quantum optics}}}}\
  (\bibinfo  {publisher} {Cambridge university press},\ \bibinfo {year}
  {1997})\BibitemShut {NoStop}%
\bibitem [{\citenamefont {Xu}\ \emph {et~al.}(2016)\citenamefont {Xu},
  \citenamefont {Yung}, \citenamefont {Xu}, \citenamefont {Tang}, \citenamefont
  {Li},\ and\ \citenamefont {Guo}}]{Xu2013}%
  \BibitemOpen
  \bibfield  {author} {\bibinfo {author} {\bibfnamefont {J.-S.}\ \bibnamefont
  {Xu}}, \bibinfo {author} {\bibfnamefont {M.-H.}\ \bibnamefont {Yung}},
  \bibinfo {author} {\bibfnamefont {X.-Y.}\ \bibnamefont {Xu}}, \bibinfo
  {author} {\bibfnamefont {J.-S.}\ \bibnamefont {Tang}}, \bibinfo {author}
  {\bibfnamefont {C.-F.}\ \bibnamefont {Li}}, \ and\ \bibinfo {author}
  {\bibfnamefont {G.-C.}\ \bibnamefont {Guo}},\ }\href {\doibase
  10.1126/sciadv.1500672} {\bibfield  {journal} {\bibinfo  {journal} {Sci.
  Adv.}\ }\textbf {\bibinfo {volume} {2}},\ \bibinfo {pages} {e1500672}
  (\bibinfo {year} {2016})}\BibitemShut {NoStop}%
\bibitem [{\citenamefont {Tillmann}\ \emph {et~al.}(2015)\citenamefont
  {Tillmann}, \citenamefont {Tan}, \citenamefont {Stoeckl}, \citenamefont
  {Sanders}, \citenamefont {de~Guise}, \citenamefont {Heilmann}, \citenamefont
  {Nolte}, \citenamefont {Szameit},\ and\ \citenamefont
  {Walther}}]{Tillmann2015}%
  \BibitemOpen
  \bibfield  {author} {\bibinfo {author} {\bibfnamefont {M.}~\bibnamefont
  {Tillmann}}, \bibinfo {author} {\bibfnamefont {S.-H.}\ \bibnamefont {Tan}},
  \bibinfo {author} {\bibfnamefont {S.~E.}\ \bibnamefont {Stoeckl}}, \bibinfo
  {author} {\bibfnamefont {B.~C.}\ \bibnamefont {Sanders}}, \bibinfo {author}
  {\bibfnamefont {H.}~\bibnamefont {de~Guise}}, \bibinfo {author}
  {\bibfnamefont {R.}~\bibnamefont {Heilmann}}, \bibinfo {author}
  {\bibfnamefont {S.}~\bibnamefont {Nolte}}, \bibinfo {author} {\bibfnamefont
  {A.}~\bibnamefont {Szameit}}, \ and\ \bibinfo {author} {\bibfnamefont
  {P.}~\bibnamefont {Walther}},\ }\href {\doibase 10.1103/PhysRevX.5.041015}
  {\bibfield  {journal} {\bibinfo  {journal} {Phys. Rev. X}\ }\textbf {\bibinfo
  {volume} {5}},\ \bibinfo {pages} {041015} (\bibinfo {year}
  {2015})}\BibitemShut {NoStop}%
\bibitem [{\citenamefont {Biggerstaff}\ \emph {et~al.}(2016)\citenamefont
  {Biggerstaff}, \citenamefont {Heilmann}, \citenamefont {Zecevik},
  \citenamefont {Gr{\"{a}}fe}, \citenamefont {Broome}, \citenamefont
  {Fedrizzi}, \citenamefont {Nolte}, \citenamefont {Szameit}, \citenamefont
  {White},\ and\ \citenamefont {Kassal}}]{Biggerstaff2016}%
  \BibitemOpen
  \bibfield  {author} {\bibinfo {author} {\bibfnamefont {D.~N.}\ \bibnamefont
  {Biggerstaff}}, \bibinfo {author} {\bibfnamefont {R.}~\bibnamefont
  {Heilmann}}, \bibinfo {author} {\bibfnamefont {A.~a.}\ \bibnamefont
  {Zecevik}}, \bibinfo {author} {\bibfnamefont {M.}~\bibnamefont
  {Gr{\"{a}}fe}}, \bibinfo {author} {\bibfnamefont {M.~a.}\ \bibnamefont
  {Broome}}, \bibinfo {author} {\bibfnamefont {A.}~\bibnamefont {Fedrizzi}},
  \bibinfo {author} {\bibfnamefont {S.}~\bibnamefont {Nolte}}, \bibinfo
  {author} {\bibfnamefont {A.}~\bibnamefont {Szameit}}, \bibinfo {author}
  {\bibfnamefont {A.~G.}\ \bibnamefont {White}}, \ and\ \bibinfo {author}
  {\bibfnamefont {I.}~\bibnamefont {Kassal}},\ }\href {\doibase
  10.1038/ncomms11282} {\bibfield  {journal} {\bibinfo  {journal} {Nat.
  Commun.}\ }\textbf {\bibinfo {volume} {7}},\ \bibinfo {pages} {11282}
  (\bibinfo {year} {2016})}\BibitemShut {NoStop}%
\bibitem [{\citenamefont {Reck}\ \emph {et~al.}(1994)\citenamefont {Reck},
  \citenamefont {Zeilinger}, \citenamefont {Bernstein},\ and\ \citenamefont
  {Bertani}}]{Reck1994}%
  \BibitemOpen
  \bibfield  {author} {\bibinfo {author} {\bibfnamefont {M.}~\bibnamefont
  {Reck}}, \bibinfo {author} {\bibfnamefont {A.}~\bibnamefont {Zeilinger}},
  \bibinfo {author} {\bibfnamefont {H.~J.}\ \bibnamefont {Bernstein}}, \ and\
  \bibinfo {author} {\bibfnamefont {P.}~\bibnamefont {Bertani}},\ }\href
  {\doibase 10.1103/PhysRevLett.73.58} {\bibfield  {journal} {\bibinfo
  {journal} {Phys. Rev. Lett.}\ }\textbf {\bibinfo {volume} {73}},\ \bibinfo
  {pages} {58} (\bibinfo {year} {1994})}\BibitemShut {NoStop}%
\bibitem [{\citenamefont {Crespi}\ \emph {et~al.}(2011)\citenamefont {Crespi},
  \citenamefont {Ramponi}, \citenamefont {Osellame}, \citenamefont {Sansoni},
  \citenamefont {Bongioanni}, \citenamefont {Sciarrino}, \citenamefont
  {Vallone},\ and\ \citenamefont {Mataloni}}]{Crespi2011}%
  \BibitemOpen
  \bibfield  {author} {\bibinfo {author} {\bibfnamefont {A.}~\bibnamefont
  {Crespi}}, \bibinfo {author} {\bibfnamefont {R.}~\bibnamefont {Ramponi}},
  \bibinfo {author} {\bibfnamefont {R.}~\bibnamefont {Osellame}}, \bibinfo
  {author} {\bibfnamefont {L.}~\bibnamefont {Sansoni}}, \bibinfo {author}
  {\bibfnamefont {I.}~\bibnamefont {Bongioanni}}, \bibinfo {author}
  {\bibfnamefont {F.}~\bibnamefont {Sciarrino}}, \bibinfo {author}
  {\bibfnamefont {G.}~\bibnamefont {Vallone}}, \ and\ \bibinfo {author}
  {\bibfnamefont {P.}~\bibnamefont {Mataloni}},\ }\href {\doibase
  10.1038/ncomms1570} {\bibfield  {journal} {\bibinfo  {journal} {Nat.
  Commun.}\ }\textbf {\bibinfo {volume} {2}},\ \bibinfo {pages} {566} (\bibinfo
  {year} {2011})}\BibitemShut {NoStop}%
\bibitem [{\citenamefont {Shadbolt}\ \emph {et~al.}(2011)\citenamefont
  {Shadbolt}, \citenamefont {Verde}, \citenamefont {Peruzzo}, \citenamefont
  {Politi}, \citenamefont {Laing}, \citenamefont {Lobino}, \citenamefont
  {Matthews}, \citenamefont {Thompson},\ and\ \citenamefont
  {O'Brien}}]{Shadbolt2012}%
  \BibitemOpen
  \bibfield  {author} {\bibinfo {author} {\bibfnamefont {P.~J.}\ \bibnamefont
  {Shadbolt}}, \bibinfo {author} {\bibfnamefont {M.~R.}\ \bibnamefont {Verde}},
  \bibinfo {author} {\bibfnamefont {A.}~\bibnamefont {Peruzzo}}, \bibinfo
  {author} {\bibfnamefont {A.}~\bibnamefont {Politi}}, \bibinfo {author}
  {\bibfnamefont {A.}~\bibnamefont {Laing}}, \bibinfo {author} {\bibfnamefont
  {M.}~\bibnamefont {Lobino}}, \bibinfo {author} {\bibfnamefont {J.~C.~F.}\
  \bibnamefont {Matthews}}, \bibinfo {author} {\bibfnamefont {M.~G.}\
  \bibnamefont {Thompson}}, \ and\ \bibinfo {author} {\bibfnamefont {J.~L.}\
  \bibnamefont {O'Brien}},\ }\href {\doibase 10.1038/nphoton.2011.283}
  {\bibfield  {journal} {\bibinfo  {journal} {Nat. Photonics}\ }\textbf
  {\bibinfo {volume} {6}},\ \bibinfo {pages} {45} (\bibinfo {year}
  {2011})}\BibitemShut {NoStop}%
\bibitem [{\citenamefont {Carolan}\ \emph {et~al.}(2015)\citenamefont
  {Carolan}, \citenamefont {Harrold}, \citenamefont {Sparrow}, \citenamefont
  {Martin-Lopez}, \citenamefont {Russell}, \citenamefont {Silverstone},
  \citenamefont {Shadbolt}, \citenamefont {Matsuda}, \citenamefont {Oguma},
  \citenamefont {Itoh}, \citenamefont {Marshall}, \citenamefont {Thompson},
  \citenamefont {Matthews}, \citenamefont {Hashimoto}, \citenamefont
  {O'Brien},\ and\ \citenamefont {Laing}}]{Carolan2015}%
  \BibitemOpen
  \bibfield  {author} {\bibinfo {author} {\bibfnamefont {J.}~\bibnamefont
  {Carolan}}, \bibinfo {author} {\bibfnamefont {C.}~\bibnamefont {Harrold}},
  \bibinfo {author} {\bibfnamefont {C.}~\bibnamefont {Sparrow}}, \bibinfo
  {author} {\bibfnamefont {E.}~\bibnamefont {Martin-Lopez}}, \bibinfo {author}
  {\bibfnamefont {N.~J.}\ \bibnamefont {Russell}}, \bibinfo {author}
  {\bibfnamefont {J.~W.}\ \bibnamefont {Silverstone}}, \bibinfo {author}
  {\bibfnamefont {P.~J.}\ \bibnamefont {Shadbolt}}, \bibinfo {author}
  {\bibfnamefont {N.}~\bibnamefont {Matsuda}}, \bibinfo {author} {\bibfnamefont
  {M.}~\bibnamefont {Oguma}}, \bibinfo {author} {\bibfnamefont
  {M.}~\bibnamefont {Itoh}}, \bibinfo {author} {\bibfnamefont {G.~D.}\
  \bibnamefont {Marshall}}, \bibinfo {author} {\bibfnamefont {M.~G.}\
  \bibnamefont {Thompson}}, \bibinfo {author} {\bibfnamefont {J.~C.~F.}\
  \bibnamefont {Matthews}}, \bibinfo {author} {\bibfnamefont {T.}~\bibnamefont
  {Hashimoto}}, \bibinfo {author} {\bibfnamefont {J.~L.}\ \bibnamefont
  {O'Brien}}, \ and\ \bibinfo {author} {\bibfnamefont {A.}~\bibnamefont
  {Laing}},\ }\href {\doibase 10.1126/science.aab3642} {\bibfield  {journal}
  {\bibinfo  {journal} {Science}\ }\textbf {\bibinfo {volume} {349}},\ \bibinfo
  {pages} {711} (\bibinfo {year} {2015})}\BibitemShut {NoStop}%
\bibitem [{\citenamefont {Aaronson}\ and\ \citenamefont
  {Arkhipov}(2013)}]{Aaronson2013a}%
  \BibitemOpen
  \bibfield  {author} {\bibinfo {author} {\bibfnamefont {S.}~\bibnamefont
  {Aaronson}}\ and\ \bibinfo {author} {\bibfnamefont {A.}~\bibnamefont
  {Arkhipov}},\ }\href {\doibase 10.4086/toc.2013.v009a004} {\bibfield
  {journal} {\bibinfo  {journal} {Theory Comput.}\ }\textbf {\bibinfo {volume}
  {9}},\ \bibinfo {pages} {143} (\bibinfo {year} {2013})}\BibitemShut {NoStop}%
\bibitem [{\citenamefont {Broome}\ \emph {et~al.}(2013)\citenamefont {Broome},
  \citenamefont {Fedrizzi}, \citenamefont {Rahimi-Keshari}, \citenamefont
  {Dove}, \citenamefont {Aaronson}, \citenamefont {Ralph},\ and\ \citenamefont
  {White}}]{Broome2013}%
  \BibitemOpen
  \bibfield  {author} {\bibinfo {author} {\bibfnamefont {M.~A.}\ \bibnamefont
  {Broome}}, \bibinfo {author} {\bibfnamefont {A.}~\bibnamefont {Fedrizzi}},
  \bibinfo {author} {\bibfnamefont {S.}~\bibnamefont {Rahimi-Keshari}},
  \bibinfo {author} {\bibfnamefont {J.}~\bibnamefont {Dove}}, \bibinfo {author}
  {\bibfnamefont {S.}~\bibnamefont {Aaronson}}, \bibinfo {author}
  {\bibfnamefont {T.~C.}\ \bibnamefont {Ralph}}, \ and\ \bibinfo {author}
  {\bibfnamefont {A.~G.}\ \bibnamefont {White}},\ }\href {\doibase
  10.1126/science.1231440} {\bibfield  {journal} {\bibinfo  {journal} {Science
  (80-. ).}\ }\textbf {\bibinfo {volume} {339}},\ \bibinfo {pages} {794}
  (\bibinfo {year} {2013})}\BibitemShut {NoStop}%
\bibitem [{\citenamefont {Spring}\ \emph {et~al.}(2013)\citenamefont {Spring},
  \citenamefont {Metcalf}, \citenamefont {Humphreys}, \citenamefont
  {Kolthammer}, \citenamefont {Jin}, \citenamefont {Barbieri}, \citenamefont
  {Datta}, \citenamefont {Thomas-Peter}, \citenamefont {Langford},
  \citenamefont {Kundys}, \citenamefont {Gates}, \citenamefont {Smith},
  \citenamefont {Smith},\ and\ \citenamefont {Walmsley}}]{Spring2013}%
  \BibitemOpen
  \bibfield  {author} {\bibinfo {author} {\bibfnamefont {J.~B.}\ \bibnamefont
  {Spring}}, \bibinfo {author} {\bibfnamefont {B.~J.}\ \bibnamefont {Metcalf}},
  \bibinfo {author} {\bibfnamefont {P.~C.}\ \bibnamefont {Humphreys}}, \bibinfo
  {author} {\bibfnamefont {W.~S.}\ \bibnamefont {Kolthammer}}, \bibinfo
  {author} {\bibfnamefont {X.-M.}\ \bibnamefont {Jin}}, \bibinfo {author}
  {\bibfnamefont {M.}~\bibnamefont {Barbieri}}, \bibinfo {author}
  {\bibfnamefont {A.}~\bibnamefont {Datta}}, \bibinfo {author} {\bibfnamefont
  {N.}~\bibnamefont {Thomas-Peter}}, \bibinfo {author} {\bibfnamefont {N.~K.}\
  \bibnamefont {Langford}}, \bibinfo {author} {\bibfnamefont {D.}~\bibnamefont
  {Kundys}}, \bibinfo {author} {\bibfnamefont {J.~C.}\ \bibnamefont {Gates}},
  \bibinfo {author} {\bibfnamefont {B.~J.}\ \bibnamefont {Smith}}, \bibinfo
  {author} {\bibfnamefont {P.~G.~R.}\ \bibnamefont {Smith}}, \ and\ \bibinfo
  {author} {\bibfnamefont {I.~A.}\ \bibnamefont {Walmsley}},\ }\href {\doibase
  10.1126/science.1231692} {\bibfield  {journal} {\bibinfo  {journal}
  {Science}\ }\textbf {\bibinfo {volume} {339}},\ \bibinfo {pages} {798}
  (\bibinfo {year} {2013})}\BibitemShut {NoStop}%
\bibitem [{\citenamefont {Tillmann}\ \emph {et~al.}(2013)\citenamefont
  {Tillmann}, \citenamefont {Daki{\'{c}}}, \citenamefont {Heilmann},
  \citenamefont {Nolte}, \citenamefont {Szameit},\ and\ \citenamefont
  {Walther}}]{Tillmann2013}%
  \BibitemOpen
  \bibfield  {author} {\bibinfo {author} {\bibfnamefont {M.}~\bibnamefont
  {Tillmann}}, \bibinfo {author} {\bibfnamefont {B.}~\bibnamefont
  {Daki{\'{c}}}}, \bibinfo {author} {\bibfnamefont {R.}~\bibnamefont
  {Heilmann}}, \bibinfo {author} {\bibfnamefont {S.}~\bibnamefont {Nolte}},
  \bibinfo {author} {\bibfnamefont {A.}~\bibnamefont {Szameit}}, \ and\
  \bibinfo {author} {\bibfnamefont {P.}~\bibnamefont {Walther}},\ }\href
  {\doibase 10.1038/nphoton.2013.102} {\bibfield  {journal} {\bibinfo
  {journal} {Nat. Photonics}\ }\textbf {\bibinfo {volume} {7}},\ \bibinfo
  {pages} {540} (\bibinfo {year} {2013})}\BibitemShut {NoStop}%
\bibitem [{\citenamefont {Crespi}\ \emph {et~al.}(2013)\citenamefont {Crespi},
  \citenamefont {Osellame}, \citenamefont {Ramponi}, \citenamefont {Brod},
  \citenamefont {Galv{\~{a}}o}, \citenamefont {Spagnolo}, \citenamefont
  {Vitelli}, \citenamefont {Maiorino}, \citenamefont {Mataloni},\ and\
  \citenamefont {Sciarrino}}]{Crespi2013}%
  \BibitemOpen
  \bibfield  {author} {\bibinfo {author} {\bibfnamefont {A.}~\bibnamefont
  {Crespi}}, \bibinfo {author} {\bibfnamefont {R.}~\bibnamefont {Osellame}},
  \bibinfo {author} {\bibfnamefont {R.}~\bibnamefont {Ramponi}}, \bibinfo
  {author} {\bibfnamefont {D.~J.}\ \bibnamefont {Brod}}, \bibinfo {author}
  {\bibfnamefont {E.~F.}\ \bibnamefont {Galv{\~{a}}o}}, \bibinfo {author}
  {\bibfnamefont {N.}~\bibnamefont {Spagnolo}}, \bibinfo {author}
  {\bibfnamefont {C.}~\bibnamefont {Vitelli}}, \bibinfo {author} {\bibfnamefont
  {E.}~\bibnamefont {Maiorino}}, \bibinfo {author} {\bibfnamefont
  {P.}~\bibnamefont {Mataloni}}, \ and\ \bibinfo {author} {\bibfnamefont
  {F.}~\bibnamefont {Sciarrino}},\ }\href {\doibase 10.1038/nphoton.2013.112}
  {\bibfield  {journal} {\bibinfo  {journal} {Nat. Photonics}\ }\textbf
  {\bibinfo {volume} {7}},\ \bibinfo {pages} {545} (\bibinfo {year}
  {2013})}\BibitemShut {NoStop}%
\bibitem [{\citenamefont {Spagnolo}\ \emph {et~al.}(2014)\citenamefont
  {Spagnolo}, \citenamefont {Vitelli}, \citenamefont {Bentivegna},
  \citenamefont {Brod}, \citenamefont {Crespi}, \citenamefont {Flamini},
  \citenamefont {Giacomini}, \citenamefont {Milani}, \citenamefont {Ramponi},
  \citenamefont {Mataloni}, \citenamefont {Osellame}, \citenamefont
  {Galv{\~{a}}o},\ and\ \citenamefont {Sciarrino}}]{Spagnolo2014}%
  \BibitemOpen
  \bibfield  {author} {\bibinfo {author} {\bibfnamefont {N.}~\bibnamefont
  {Spagnolo}}, \bibinfo {author} {\bibfnamefont {C.}~\bibnamefont {Vitelli}},
  \bibinfo {author} {\bibfnamefont {M.}~\bibnamefont {Bentivegna}}, \bibinfo
  {author} {\bibfnamefont {D.~J.}\ \bibnamefont {Brod}}, \bibinfo {author}
  {\bibfnamefont {A.}~\bibnamefont {Crespi}}, \bibinfo {author} {\bibfnamefont
  {F.}~\bibnamefont {Flamini}}, \bibinfo {author} {\bibfnamefont
  {S.}~\bibnamefont {Giacomini}}, \bibinfo {author} {\bibfnamefont
  {G.}~\bibnamefont {Milani}}, \bibinfo {author} {\bibfnamefont
  {R.}~\bibnamefont {Ramponi}}, \bibinfo {author} {\bibfnamefont
  {P.}~\bibnamefont {Mataloni}}, \bibinfo {author} {\bibfnamefont
  {R.}~\bibnamefont {Osellame}}, \bibinfo {author} {\bibfnamefont {E.~F.}\
  \bibnamefont {Galv{\~{a}}o}}, \ and\ \bibinfo {author} {\bibfnamefont
  {F.}~\bibnamefont {Sciarrino}},\ }\href {\doibase 10.1038/nphoton.2014.135}
  {\bibfield  {journal} {\bibinfo  {journal} {Nat. Photonics}\ }\textbf
  {\bibinfo {volume} {8}},\ \bibinfo {pages} {615} (\bibinfo {year}
  {2014})}\BibitemShut {NoStop}%
\bibitem [{\citenamefont {Bentivegna}\ \emph {et~al.}(2015)\citenamefont
  {Bentivegna}, \citenamefont {Spagnolo}, \citenamefont {Vitelli},
  \citenamefont {Flamini}, \citenamefont {Viggianiello}, \citenamefont
  {Latmiral}, \citenamefont {Mataloni}, \citenamefont {Brod}, \citenamefont
  {Galvao}, \citenamefont {Crespi}, \citenamefont {Ramponi}, \citenamefont
  {Osellame},\ and\ \citenamefont {Sciarrino}}]{Bentivegna2015}%
  \BibitemOpen
  \bibfield  {author} {\bibinfo {author} {\bibfnamefont {M.}~\bibnamefont
  {Bentivegna}}, \bibinfo {author} {\bibfnamefont {N.}~\bibnamefont
  {Spagnolo}}, \bibinfo {author} {\bibfnamefont {C.}~\bibnamefont {Vitelli}},
  \bibinfo {author} {\bibfnamefont {F.}~\bibnamefont {Flamini}}, \bibinfo
  {author} {\bibfnamefont {N.}~\bibnamefont {Viggianiello}}, \bibinfo {author}
  {\bibfnamefont {L.}~\bibnamefont {Latmiral}}, \bibinfo {author}
  {\bibfnamefont {P.}~\bibnamefont {Mataloni}}, \bibinfo {author}
  {\bibfnamefont {D.~J.}\ \bibnamefont {Brod}}, \bibinfo {author}
  {\bibfnamefont {E.~F.}\ \bibnamefont {Galvao}}, \bibinfo {author}
  {\bibfnamefont {A.}~\bibnamefont {Crespi}}, \bibinfo {author} {\bibfnamefont
  {R.}~\bibnamefont {Ramponi}}, \bibinfo {author} {\bibfnamefont
  {R.}~\bibnamefont {Osellame}}, \ and\ \bibinfo {author} {\bibfnamefont
  {F.}~\bibnamefont {Sciarrino}},\ }\href {\doibase 10.1126/sciadv.1400255}
  {\bibfield  {journal} {\bibinfo  {journal} {Sci. Adv.}\ }\textbf {\bibinfo
  {volume} {1}},\ \bibinfo {pages} {e1400255} (\bibinfo {year}
  {2015})}\BibitemShut {NoStop}%
\bibitem [{\citenamefont {Aaronson}\ and\ \citenamefont
  {Hance}(2012)}]{Aaronson2012b}%
  \BibitemOpen
  \bibfield  {author} {\bibinfo {author} {\bibfnamefont {S.}~\bibnamefont
  {Aaronson}}\ and\ \bibinfo {author} {\bibfnamefont {T.}~\bibnamefont
  {Hance}},\ }\href {http://www.rintonpress.com/journals/qiconline.html}
  {\bibfield  {journal} {\bibinfo  {journal} {Quantum Inf. Comput.}\ }\textbf
  {\bibinfo {volume} {14}},\ \bibinfo {pages} {541} (\bibinfo {year}
  {2012})}\BibitemShut {NoStop}%
\bibitem [{\citenamefont {Ryser}(1963)}]{Ryser}%
  \BibitemOpen
  \bibfield  {author} {\bibinfo {author} {\bibfnamefont {H.~J.}\ \bibnamefont
  {Ryser}},\ }\href@noop {} {\emph {\bibinfo {title} {{Combinatorial
  mathematics}}}}\ (\bibinfo  {publisher} {Mathematical Association of
  America},\ \bibinfo {year} {1963})\BibitemShut {NoStop}%
\bibitem [{\citenamefont {Glynn}(2010)}]{Glynn2010}%
  \BibitemOpen
  \bibfield  {author} {\bibinfo {author} {\bibfnamefont {D.~G.}\ \bibnamefont
  {Glynn}},\ }\href {\doibase 10.1016/j.ejc.2010.01.010} {\bibfield  {journal}
  {\bibinfo  {journal} {Eur. J. Comb.}\ }\textbf {\bibinfo {volume} {31}},\
  \bibinfo {pages} {1887} (\bibinfo {year} {2010})}\BibitemShut {NoStop}%
\bibitem [{\citenamefont {Glynn}(2013)}]{Glynn2013}%
  \BibitemOpen
  \bibfield  {author} {\bibinfo {author} {\bibfnamefont {D.~G.}\ \bibnamefont
  {Glynn}},\ }\href@noop {} {\bibfield  {journal} {\bibinfo  {journal}
  {Designs, Codes, and Cryptography}\ }\textbf {\bibinfo {volume} {68}},\
  \bibinfo {pages} {39} (\bibinfo {year} {2013})}\BibitemShut {NoStop}%
\bibitem [{\citenamefont {Gurvits}(2005)}]{Gurvits2005}%
  \BibitemOpen
  \bibfield  {author} {\bibinfo {author} {\bibfnamefont {L.}~\bibnamefont
  {Gurvits}},\ }in\ \href {\doibase 10.1007/11549345_39} {\emph {\bibinfo
  {booktitle} {Math. Found. Comput. Sci. 2005}}}\ (\bibinfo {year} {2005})\
  pp.\ \bibinfo {pages} {447--458}\BibitemShut {NoStop}%
\bibitem [{\citenamefont {Huh}(2016)}]{huh2016}%
  \BibitemOpen
  \bibfield  {author} {\bibinfo {author} {\bibfnamefont {J.}~\bibnamefont
  {Huh}},\ }\href@noop {} {\bibfield  {journal} {\bibinfo  {journal}
  {arXiv:1605.08506v1}\ } (\bibinfo {year} {2016})}\BibitemShut {NoStop}%
\bibitem [{\citenamefont {Caianiello}(1953)}]{Caianiello1953}%
  \BibitemOpen
  \bibfield  {author} {\bibinfo {author} {\bibfnamefont {E.~R.}\ \bibnamefont
  {Caianiello}},\ }\href {\doibase 10.1007/BF02781659} {\bibfield  {journal}
  {\bibinfo  {journal} {Nuovo Cim.}\ }\textbf {\bibinfo {volume} {10}},\
  \bibinfo {pages} {1634} (\bibinfo {year} {1953})}\BibitemShut {NoStop}%
\end{thebibliography}

%

\appendix

\section{Appendix: Validity of the optimal strategies} Here we show that the upper bounds of the two examples can be saturated with the strategies discussed in the main text. First, consider putting $n$ bosons from two modes, $a_1^\dagger$ and $a_2^\dagger$, into $2n$ bosons in one mode. For a rotational operator $U$, where 
\begin{align}
  Ua_1^\dag {U^\dag } = \cos \theta a_1^\dag  + \sin \theta a_2^\dag \ , \label{App_Ua_1} \\
  Ua_2^\dag {U^\dag } = \cos \theta a_2^\dag  - \sin \theta a_1^\dag \ .   \label{App_Ua_2}
\end{align}
The corresponding transition amplitude $A\left( {2n,0|n,n} \right)$ is given by,
\begin{equation}
A\left( {2n,0|n,n} \right) = \frac{{\left\langle {{\text{vac}}} \right|a_1^{2n} U a_1^{\dag n} a_2^{\dag n} \left| {{\text{vac}}} \right\rangle }}{{\sqrt {(2n)!n!n!} }} ,
\end{equation}
where $\left\langle {{\text{vac}}} \right|a_1^{2n}Ua_1^\dag a_2^\dag \left| {{\text{vac}}} \right\rangle  = {\sin ^n}\left( {2\theta } \right)(2n)!/{2^n}$. Therefore, the maximum probability can be achieved by setting $\theta  = \pi /4$ (i.e., a 50:50 beamsplitter), which coincides with the upper bound in the main text,
\begin{equation}
{P_{\max }} (2n,0|n,n) = {\left| A \right|^2} = \frac{{(2n)!}}{{{{(n!)}^2}{2^{2n}}}} \ .
\end{equation}

For the second question, where an extra boson is added to a mode with $n$ bosons, we choose the same form of $U$ as in Eq.~(\ref{App_Ua_1}) and Eq.~(\ref{App_Ua_2}), but with ${\sin ^2}\theta_n  = {(n + 1)^{ - 1}}$. In this way, we have 
\begin{equation}
A\left( { n +1,0|n,1} \right) = \frac{{\left\langle {{\text{vac}}} \right|a_1^{n + 1} U a_1^{\dag n} a_2^{\dag } \left| {{\text{vac}}} \right\rangle }}{{\sqrt {(n + 1)!n!} }} \ ,
\end{equation}
where $\left\langle {{\text{vac}}} \right|a_1^{n + 1} U a_1^{\dag n} a_2^{\dag }\left| {{\text{vac}}} \right\rangle  = \cos^n \theta \ {\sin }\theta (n + 1)!$. Therefore, the maximum probability is given by 
\begin{equation}
{P_{\max }}\left( {n + 1,0|n,1} \right)  = {\left( {\frac{n}{{n + 1}}} \right)^n} \ ,
\end{equation}
which coincides with the upper bound.

\section{Appendix: Chebyshev's inequality for random variables with discrete complex values}
Let us discuss more on the Chebyshev's inequality required for our bound. The majority of textbooks deals with discrete real random variables, but here we are considering discrete complex numbers. To ensure the inequality is still applicable to our case, we provide an appendix to derive it by assuming complex random variables in the beginning.

Consider the generalization of the Markov inequality for a discrete random variable $X$, which takes on discrete complex values from a set $\{ x_i \}$. The Markov's bound implies that, for a non-negative $\alpha$,
\begin{equation}
\Pr [\left| X \right| \geqslant \alpha ] \leqslant \frac{{{\mathbb E}(\left| X \right|)}}{\alpha } \ .    
\end{equation}

\hrule
\begin{proof}
By definition, the expectation value of the absolute values is given by ${\mathbb E}(\left| X \right|) = \sum\limits_a {a \cdot \Pr } (\left| X \right| = a)$,
where the sum is over all possible values of $|X|$, labeled by $a$. We can sort out the parts where $a \geq \alpha$, which means that, 
\begin{equation}
{\mathbb{E}}(\left| X \right|) \geqslant \sum\limits_{a \geqslant \alpha } {a \cdot \Pr } (\left| X \right| = a) \ .
\end{equation}
Furthermore, we also get a smaller value by replacing all $a$'s by $\alpha$, i.e., 
\begin{equation}
{\mathbb{E}}(\left| X \right|) \geqslant \alpha \sum\limits_{a \geqslant \alpha } {\Pr } (\left| X \right| = a) \ .    
\end{equation}
Lastly, we can identify the last term as the probability where $\left| X \right| \geqslant \alpha $, i.e., 
\begin{equation}
\Pr [\left| X \right| \geqslant \alpha ] = \sum\limits_{a \geqslant \alpha } {\Pr } (\left| X \right| = a) \ ,    
\end{equation}
which yields the desired result.
\end{proof}
\hrule \
\\

Now, define a variable, $\mu$, to represent the expectation value of $X$, i.e., 
\begin{equation}
\mu  \equiv {\mathbb{E}}(X) = \sum\limits_i {x_i \cdot \Pr } (X = x_i) \ .
\end{equation}
 Furthermore, we define a new random variable 
\begin{equation}
Y \equiv {\left| {X - \mu } \right|^2} = (X-\mu)(X^*-\mu^*) \ ,    
\end{equation}
where ${\mathbb{E}}(Y) = \sum\limits_i {{{\left| {x_i - \mu } \right|}^2}\Pr } ( X = x_i) $.
Note that we can also write,
\begin{equation}
{\mathbb{E}}(Y) = {\text{Var}}\left( X \right) \equiv {\text{E}}(|X{|^2}) - |\mu|^2 \ ,    
\end{equation}
and that, $\Pr (\left| {X - \mu } \right| \geqslant \alpha ) = \Pr (Y \geqslant {\alpha ^2})$.
Therefore, if we apply the Markov inequality as follows, 
\begin{equation}
\Pr (Y \geqslant {\alpha ^2}) \leqslant \frac{{{\text{E}}(Y)}}{{{\alpha ^2}}} = \frac{{{\text{Var}}(X)}}{{{\alpha ^2}}} \ ,
\end{equation}
then, we obtain the Chebyshev's inequality, 
\begin{equation}
\Pr (|X - \mu | \geqslant {\alpha}) \leqslant \frac{{{\text{Var}}(X)}}{{{\alpha ^2}}} \ .
\end{equation}

Now, let us consider $n$ independent complex random variables, $X_1$, $X_2$, ..., $X_n$, with ${\mathbb E}(X_i) = \mu_i$ and ${\rm Var}(X_i) = \sigma_i^2$. Then, the Chebyshev's inequality implies that 
\begin{equation}
\Pr (\left| {\sum\limits_{i = 1}^n {{X_i}}  - \sum\limits_{i = 1}^n {{\mu _i}} } \right| \geqslant \alpha ) \leqslant \frac{{\sum\nolimits_{i = 1}^n {\sigma _i^2} }}{{{\alpha ^2}}} \ ,
\end{equation}
where we used the fact that ${\text{Var}}(\sum\nolimits_{i = 1}^n {{X_i}} ) = \sum\nolimits_{i = 1}^n {{\text{Var}}({X_i})}$.

For identical random variables, where all $\mu_i = \mu$ and $\sigma_i = \sigma$, we have 
\begin{equation}
\Pr (\left| {\frac{{\sum\nolimits_{i = 1}^n {{X_i}} }}{n} - \mu } \right| \geqslant \epsilon ) \leqslant \frac{{{\sigma ^2}}}{{n{\epsilon ^2}}} \ ,
\end{equation}
which is the inequality needed for our results.

\end{document}